# Polarimetry, photometry, and spectroscopy of comet C/2009 P(Garradd)

Oleksandra Ivanova [a, b,\*], Vera Rosenbush[a], Viktor Afanasiev[c], Nikolai Kiselev[a, d]

[a] *Main Astronomical Observatory of National Academy of Sciences, Kyiv, Ukraine*
[b] *Astronomical Institute of the Slovak Academy of Sciences, SK-05960 Tatranská Lomnica, Slovak Republic*
[c] *Special Astrophysical Observatory, Russian Academy of Sciences, Nizhnij Arkhyz, Russia*
[d] *Crimean Astrophysical Observatory, Nauchnij, Crimea*

## ABSTRACT

We present the results of photometry, linear spectropolarimetry, and imaging circular polarimetry of comet C/2009 P1 (Garradd) performed at the 6-m telescope BTA of the Special Astrophysical Observatory (Russia) equipped by the multi-mode focal reducer SCORPIO-2. The comet was observed at two epochs post-perihelion: on February 2–14, 2012 at $r_h \sim 1.6$ au and $\alpha \sim 36°$ ; and on April 14–21, 2012 at $r_h \sim 2.2$ au and $\alpha \sim 27°$. The spatial maps of the relative intensity and circular polarization as well as the spectral distribution of linear polarization are presented. There were two features (dust and gas tails) oriented in the solar and antisolar directions on February 2 and 14 that allowed us to determine rotation period of the nucleus as $11.1 \pm 0.8$ hours. We detected emissions of $C_2$, $C_3$, CN, CH, $NH_2$ molecules as well as $CO^+$ and $H_2O^+$ ions, along with a high level of the dust continuum. On February 2, the degree of linear polarization in the continuum, within the wavelength range of 0.67–0.68 μm, was about $5 \pm 0.2\%$ in the near-nucleus region up to $\sim 6000$ km and decreased to about $3 \pm 0.2\%$ at $\sim 40,000$ km. After correction for the continuum contamination, the inherent degree of polarization in the emission band $C_2$ ($\Delta\nu = 0$) is about 3.3%. We detected a small increase of linear polarization with the wavelength with the spec- tral gradient $\Delta P/\Delta\lambda = +4 \pm 0.8\%/\mu m$ and $\Delta P/\Delta\lambda = +6.2 \pm 1.3\%/\mu m$, respectively, on February 2 and April 14. Linear polarization indicates that this dust-rich comet can be attributed to the high-$P_{max}$ comets. The left-handed (negative) circular polarization at the level approximately from $-0.06 \pm 0.02\%$ to $-0.4 \pm 0.02\%$ was observed at the distances up to $3 \times 10^4$ km from the nucleus on February 14 and April 21, respectively.

## 1. Introduction

Currently, attempts are underway to establish a possible taxonomy of comets on the basis of their composition and to link it to the place of their origin (Mumma and Charnley, 2011; A'Hearn et al., 2012; Cochran et al., 2015). A comparison of physical characteristics of short-period comets with those for long-period and new comets (in the Oort sense, see A'Hearn et al. (1995)) may elucidate which properties of comets are primordial and which are a product of subsequent evolution. Study of polarization properties of different comets may also provide classification of comets and understanding the main processes of dust formation in the protosolar nebula. In addition, polarimetric characteristics of comets may indicate differences in the properties of dust on the surface and inside of nuclei through observations of jets and different fragments of comets (Kiselev et al., 2002; Hadamcik and Levasseur-Regourd, 2016). For these purposes, we conduct systematic polarimetric observations of comets that are available for our instruments. Here we present the results for one of those comets, for comet C/2009 P1 (Garrard) (hereafter Garradd).

Comet Garradd was discovered by Gordon J. Garradd (Siding Spring Observatory, Australia) in August 2009 at a heliocentric distance of 8.7 au, as an object with an evident dust coma. The comet passed through perihelion on December 23, 2011 ($r = 1.55$ au) and was at closest approach to the Earth on March 5, 2012 ($\Delta = 1.27$ au). The comet was sufficiently bright (10.2 – 9.8 m) in order to conduct its ground-based and space-based observations.

Comet Garradd was unusually dust-rich, and its activity was complex and changed significantly over time (Bodewits et al., 2014; Feaga et al., 2014). The comet produced lots of dust and gas long before it reached the snow line. This means that its activity was caused by something other than water ice. Actually, Paganini et al. (2012) showed that comet Garradd is CO rich with a production rate ratio to water of 12–13%. Bodewits et al. and Feaga et al.

found that the water production rate in comet Garradd was highly asymmetric around perihelion. It very steep increased before perihelion following an $\sim r^{-6}$ relation, peaked at a rate of $2 \times 10^{29}$ mol/s for 200 days before perihelion, remained approximately constant for about 100 days, and then decreased at a lower rate ($\sim r^{-4}$) after perihelion. Combi et al. (2013) also concluded that the pre-perihelion activity was much larger than the post-perihelion and varied rather irregularly. Given the water production rates measured for the comet, Boissier et al. (2013) estimated upper limit of Garradd's nucleus radius in 5.6 km with an active fraction of its larger than 50%.

Dust production in comet Garradd also varied greatly with the heliocentric distance. At $r \approx 4$ au pre-perihelion, it had one of the highest dust-to-gas ratios ever observed in a comet (the value $Af\rho$, which is a measure of the dust production, was 5918 cm and log$Af\rho/Q$(OH) ratio was −24.6 cm×s/mol), but it had typical dust-to-gas ratios around perihelion at $r = 1.55$ au ($Af\rho = 3093$ cm and log$Af\rho/Q$(OH) was −25.3 cm×s/mol), i.e. a factor of 16 lower than at $r \approx 4$ au (Bodewits et al., 2014). Before perihelion, from October 28 to November 14, 2011, $r = 1.73 – 1.64$ au, $Af\rho$ varied from 900 to 5500 cm (Ivanova et al., 2014). After perihelion on March 14–April 16, 2012, when the comet moved from 1.9 au to 2.12 au from the Sun, logarithm of the dust-to-gas ratio was 3.5 – 3.4 (Schleicher, personal communication).

According to studies of the wavelength dependence of the scattered and thermally-emitted light, the derived bolometric albedo of comet Garradd was found to be slightly lower than the majority of cometary data obtained at similar phase angles (Sitko et al., 2013). For example, the reflectivity of cometary grains in comet Garradd was by 15 – 30% lower than that in comet Halley measured at the same phase angles.

Comparing water production rates derived from different observations, a strong extended source of water was separated which was continuously replenished (Bockelée-Morvan et al., 2012; Paganini et al., 2012; Combi et al., 2013; Feaga et al., 2014). Bodewits et al. (2014) separated the two sources of water: between $r = 3.0$ AU and perihelion ($r = 1.54$ AU), water was produced predominantly from ice in the coma; the second source could be $CO_2$ that drove the icy grains into the coma of comet Garradd (Bodewits et al., 2014; Decock et al., 2015; McKay et al., 2015).

The dust coma of comet Garradd was also studied by polarimetric methods. Kiselev et al. (2012) carried out aperture photoelectric polarimetric observations of the comet during seven sets: July 29, August 29, September 26 and 28, October 26, 2011 and March 24 and April 22, 2012. The phase angle varied within the range from 13.7° to 35.9°. Respectively, the degree of polarization of the comet changed from $-2.2 \pm 0.2$% to $4.9 \pm 0.2$% at these phase angles. It turns out that the degree polarization at the minimum for comet Garradd is considerably higher (in absolute value) than that for other comets, about $-1.5$% at a phase angle of $\sim 10°$ (Kiselev and Rosenbush, 2004). However, at other phase angles it is in a good agreement with that for dusty comets at the respective phase angles.

Hadamcik et al. (2014) performed polarimetry and photometry of comet Garradd during five periods: before perihelion (October 21–22, and 26, 2011) and after perihelion (January 22–25, February 18–20, and March 17–20, 2012) at the phase angles ranging from 28° to 35° After perihelion, an increase of activity was observed in intensity and polarization with very complex varying shapes of the jets (see, e.g., Fig. 2 in Hadamcik et al. 2014). Some of them were preferentially oriented in the solar direction. Depending on the distance from the nucleus and directions through the coma (up to 40,000 km), the degree of linear polarization was between $\sim$2.6% and $\sim$3% in October; between $\sim$6% and $\sim$4% in January; between $\sim$3% and $\sim$5% in February; and between $\sim$2% and $\sim$4% in March. In the outer coma, the polarization degree dropped down to 1%. Jets visible on the polarization map at distance greater than 20,000 km displayed approximately 4% polarization, whereas the surrounding polarization was 3%. The errors of measurements were within the range of 0.2–0.6%. Analyzing the activity of the comet visible on intensity images and polarization maps, the authors concluded that in large apertures the polarization of comet Garradd was similar to values obtained for other comets. Probably the comet belongs to a class of comets with a high maximum polarization (denoted as high-$P_{max}$ comets) and with the dust characterized by small submicron to micron-sized grains, possibly assembled in aggregates.

Das et al. (2013) have carried out the optical imaging polarimetry of comet Garradd on March 21 – 22, 2012 ($\alpha \approx 28°$) and May 23, 2012 ($\alpha \approx 21.6°$). In March, jets were detected. The degree of polarization varied between 3% (close to optocenter) and 0.5% in the outer coma. In May, the negative polarization in the inner coma varied between − 1.8% and − 0.2%, whereas the positive polarization was detected in the outer coma at level between 0.04% and 1.4%. Such variations of polarization in the inner and outer coma suggest that the physical properties of cometary dust vary with the distance from the optocenter.

It should be noted that the variations in linear polarization between inner and outer areas of the coma had already been discovered a long time ago, e.g., in comet 1P/Halley in remote and space observations (Dollfus and Suchail, 1987; Levasseur-Regourd et al., 1999). Actually, the physical properties of the dust changes with optocentric distance but at large distances the noise is high and the polarization has no meaning. In the inner part of the coma, the polarization may change a lot due to the nucleus rotation and the presence, for example, of jets. However, as the optocentric distance increases the integrated polarization stabilizes before decreasing when the signal/noise becomes smaller. This property is well illustrated by Hadamcik and Levasseur-Regourd (2016) on comet 73P/Schwassmann–Wachmann 3.

For several years, we have been carrying out observations of different comets with the focal reducer SCORPIO-2 mounted at the 6-m telescope of the Special Astrophysical Observatory (Russia). Some preliminary results of our long-term monitoring campaign of comet Garradd were published by Kiselev et al. (2013). These data also included the results of aperture polarimetry obtained with the 2.6 m telescope of the Crimean Astrophysical Observatory. They were mainly used for analysis of the phase-angle dependence of polarization for comet Garradd. In the present paper, the results of linear spectropolarimetry and imaging circular polarimetry are discussed along with new results of photometric and spectral observations of comet Garradd carried out on February 2–April 21, 2012. In Section 1, we present short review of relevant results of ground-based and space observations of comet Garradd, including the results of polarimetric observations. We give some basic and specific details of our observations and data reductions in Sections 2 and 3. A comprehensive analysis of data obtained and the basic results are presented in Section 4. Discussion of the results and summary are given in Sections 5 and 6, respectively.

## 2. Instrument and observations

Comet Garradd was observed during two periods: on February 2 and 14, 2012, when its heliocentric distance, $r$, increased from 1.64 to 1.70 au, geocentric distance, $\Delta$, has decreased from 1.53 to 1.39 au, and phase angle, $\alpha$, was between 36° and 35°; and on April 14 and 21, 2012, $r = 2.16 \div 2.23$, $\Delta = 1.79 \div 1.97$, and $\alpha \approx 27°$ In both periods, observations of the comet were performed at the primary focus (f/4) of the 6 m telescope BTA of the Special Astrophysical Observatory (Russia) with the multi-mode focal reducer SCORPIO-2 (Afanasiev and Moiseev, 2011; Afanasiev and Amirkhanyan, 2012; Kiselev et al., 2013). The following modes of the instrument were used for observations: direct CCD images (hereafter called Image); the low-resolution linear and circular spectropolarimetry (SpPol); and imaging circular polarimetry (ImaPol). The observations of comet Garradd were conducted in the packet mode, which allows making a sequence of exposures with different settings of the angle of rotation of the phase plate or polarization filter. In general, the number of cycles is not limited, therefore the process of measurements was iterated many times (see Table 1). The accuracy of measurements of the polarization parameters was limited only by the count statistics (Tinbergen, 1973). To improve the signal-to-noise ratio (S/N) of the measured signal, on-chip binning was applied to all observed images.

**Table 1** Log of the observations of comet Garradd.

| Date of observations, 2012, UT | $r$ (au) | $\Delta$ (au) | $\alpha$ (deg) | $PA$ (deg) | Filter/ image scale (arcsec) | $T_{exp}$(s)/N | Mode |
|---|---|---|---|---|---|---|---|
| Feb. 02.061 | 1.65 | 1.53 | 35.9 | 306.1 | VPHG940/0.36″×0.71″ | 60/12 | SpPol, $\lambda/4$, 0.35–0.85 μm, WOLL-1, binning 2 × 4 |
| Feb. 02.086 | 1.65 | 1.53 | 35.9 | 306.1 | VPHG940/0.36″×0.71″ | 30/20 | SpPol, $\lambda/2$, 0.35–0.85 μm, WOLL-1, binning 2 × 4 |
| Feb. 02.099 | 1.65 | 1.53 | 35.9 | 306.1 | g-sdss/0.36″×0.36″ | 3/1 | Image, binning 2 × 2 |
| Feb. 14.076 | 1.71 | 1.39 | 35.3 | 291.1 | RC/0.36″×0.36″ | 10/5 | Image, binning 2 × 2 |
| Feb. 14.092 | 1.70 | 1.39 | 35.4 | 291.1 | RC/0.36″×0.36″ | 30/24 | ImaPol, $\lambda/4$, WOLL-1, binning 2 × 2 |
| Apr. 14.843 | 2.16 | 1.79 | 27.4 | 112.8 | V/0.36″×0.36″ | 4/1 | Image, binning 2 × 2 |
| Apr. 14.864 | 2.17 | 1.80 | 27.4 | 112.8 | VPHG1200/0.36″×0.71″ | 120/16 | SpPol, $\lambda/2$, WOLL-1, 0.36–0.71 μm, binning 2 × 4 |
| Apr. 21.795 | 2.23 | 1.96 | 26.8 | 108.2 | RC/0.36″×0.36″ | 60/14 | ImaPol, $\lambda/4$, WOLL-1, binning 2 × 2 |

The Wollaston prism, designated WOLL-1, was used as the polarization analyzer for measurements of polarization in comet Garradd. The WOLL-1 has the angle of divergence of the ordinary, $I_o(\lambda)$, and extraordinary, $I_e(\lambda)$, rays 5° what corresponds to the operating slit height of about 2′ on the celestial sphere. We used the mask, with height 2′ and diameter 4′, which isolated non-overlapping area. Behind the slit, in the focal plane, the rotating super-achromatic plate with $\lambda/2$ or $\lambda/4$ phase is installed. For fixed positions of the $\lambda/2$ phase plate 0°, 45°, 22.5°, and 67.5°, a series of pairs of spectra $I_o(\lambda)$ and $I_e(\lambda)$ in mutually perpendicular polarization planes were obtained at the exit of the spectrograph. The slit height was 6′, and its width was changeable in the range 0.5–22″. The calibration system, consisting of a calibration lamp with He-Ne-Ar filling for wavelength calibration and halogen lamp with continuous spectrum, allowed us to correct calibration of the wavelengths and of the flat-field.

For the measurements of circular polarization, the $\lambda/4$ phase plate was inserted in the beam, which was oriented in such a way that the direction of the fast axis made the angle 45° with the direction of the main axis of the Wollaston prism. The plate rotated by two fixed angles, 0° and 90°. This measurement technique of circular polarization by SCORPIO-2 allows one to eliminate the phase shift errors of the phase plate, but requires a stable atmosphere. Since the achromatic phase plates have anisotropic optical multilayer coatings (layers), their parameters depend significantly on the convergence of the beam in which they are installed. Installation of the phase plates in a parallel beam leads to a phase shift, which depends on the distance from the center of the field of view, and ultimately to an emergence of the instrumental polarization, which depends on the position of the object in the field of view of spectropolarimeter. Due to this, the instrumental polarization can reach a value up to 2% that is a significant value in measuring small quantities (of the order of 0.05 – 0.1%). Therefore SCORPIO-2 is designed to compensate this defect, namely, a phase plate is mounted in a convergent (f/4) beam.

On February 2 and 14 and April 14, 2012, the direct photometric images of comet Garradd were obtained with the broad-band g ($\lambda_0 = 0.465/0.050$ μm) and V ($\lambda_0 = 0.551/0.088$ μm) filters of the SDSS and Johnson-Cousins photometric systems, respectively, and the narrow-band cometary continuum filter RC ($\lambda_0 = 0.684/0.009$ μm). For obtaining direct images, the large-format CCD EEV 42–90 matrix of 2048 × 2048 pixels was used as a detector. A full view of the detector was 6.1′×6.1′ with an image scale 0.18 arcsec/px. The telescope was tracked on the comet to compensate its apparent movement during the exposures. Observations of the twilight sky through the used filters were also performed to provide flat-field corrections.

Linear spectropolarimetry of comet Garradd was conducted in the long-slit mode with spectral gratings VPHG940 (February 2, 2012) and VPHG1200 (April 14, 2012) which operate in the wavelength ranges 0.35–0.85 μm and 0.36–0.71 μm, respectively. However, due to a low quantum efficiency of the grating and the radiation detector in the blue spectral region, the operating wavelength range was about 0.38–0.80 μm for the first date and 0.38–0.71 μm for the second one. The light beam from the comet was registered in spectra with the slit height of 6′ and width of 3″. The spectral resolution of the obtained spectra amounted to about 0.0005 μm. The spectrograph slit was placed such that the comet nucleus was centered in the slit and was oriented at a position angle of 111.3° and 6.7°, i.e. along the comet heliocentric velocity vector, on February 2 and April 14, respectively. The exposure for each measurement was chosen to allow getting required signal-to-noise ratio and to minimize depolarizing effect of the Earth's atmosphere (depolarization). Image of the spectra of a continuous-spectrum lamp was used for correction of flat field, which was obtained for each angle of the phase plate.

The log of observations is given in Table 1. The date of observation (the mid-cycle time), the heliocentric and geocentric distances, the phase angle, the position angle of the scattering plane ($PA$), the filter and image scale, the total exposure and number of cycles ($N$), and the mode of observations are listed in the table.

## 3. Data processing

The primary reductions of the observational data were performed using standard technique of the image processing, including dividing the science frames by flat fields after bias subtraction. The sky background was estimated in a region outside of the coma and free of faint stars. The nights were photometric and the seeing was stable around 1.5″–1.7″. However, there are some specific features for each mode of the device SCORPIO-2 which will be described below. Since measurements were differential, and, hence, we compared differential values in each image point, even very small differences between the images obtained at different angles of the phase plates, could impair the accuracy of polarization measurements. The same applied to the removal of the traces of cosmic rays and different techniques of smoothing or using the optimal aperture photometry, since all of these algorithms (procedures) could influence the appearance of an artificial instrumental polarization, which could shift the statistic estimates, and therefore contribute some depolarization effects to the final result. The achromatic phase plate is mounted near the focal plane of the telescope in the divergent beam, and therefore did not introduce any significant variations of instrumental polarization across the field.

### 3.1. Intensity images

The standard procedures of bias subtraction and flat-fielding correction were applied to all the raw data. The frames with twi-light were obtained to create the averaged flat field image. The level of sky background for each individual frame was estimated from those parts of the frame, which were not covered by the cometary coma and free of faint stars, using procedure of building histogram of counts in the image. The count corresponding to the maximum was chosen as the background sky level which was subtracted from the image. To increase the S/N ratio, the individual frames were stacked together and summed using a robust averaging method (Rousseeuw and Bassett, 1990). We used the robust estimation because it is more stable in respect to random errors and allows us to calculate an unbiased average of the measured values (Maronna et al., 2006).

Since measurements were differential, we combined the frames of the comet using only central contour of relative intensity (isophotes) which was closest to the maximum of brightness of the comet. Removal of the traces of cosmic rays was done at the final stage of the reduction via a robust parameter estimates for reducing a bias caused by outliers (Fujisawa, 2013). The flux from the comet was determined by a simple summation.

In order to segregate low-contrast structures in the images, we used an enhancement technique (Samarasinha and Larson, 2014): a rotational gradient method (Larson and Sekanina, 1984), unsharping mask, and Gauss blurring. To exclude spurious features when interpreting the obtained images, each of the digital filters was used separately for each particular image. Earlier this technique was used to pick out structures in several comets with good results (Manzini et al., 2007; Ivanova et al., 2009). The results obtained will be discussed in Section 4.1.

### 3.2. Circular polarization images

For the measurement of circular polarization in the comet, the direct images of the comet in ordinary $I_o(\lambda)$ and extraordinary $I_e(\lambda)$ rays were obtained in the cometary continuum filter RC. Several cycles of measurements were carried out during the observations, when we successively recorded the image for pairs of angles (0°, 90°) for the $\lambda/4$ plate. The systematic changes in the relative depolarization at different phase plate angles exceeding the statistical variations were interpreted as variations in the polarization channel sensitivity and were taken into account in the calculations of circular polarization. The errors, which were introduced by the influence of the atmosphere, because the measurements were obtained not simultaneously, were taken good enough into account in the processing which is described in detail by Afanasiev and Amirkhanyan (2012). The degree of circular polarization in terms of Stokes parameter (see, e.g., Gehrels (1974), Mishchenko et al. (2002), Clarke (2010)), is defined as $P_{circ}=V/I$, where the measured Stokes parameters $V$ normalized to the integral intensity $I$ were calculated according to the relation from Afanasiev and Amirkhanyan (2012):

$$\frac{V}{I} = 0.5\left[\frac{I(\lambda)_0 - I(\lambda)_{90}}{I(\lambda)_0 + I(\lambda)_{90}}\right] - 0.5\left[\frac{I(\lambda)_{45} - I(\lambda)_{135}}{I(\lambda)_{45} + I(\lambda)_{135}}\right],$$

where $I(\lambda)_0$, $I(\lambda)_{90}$, $I(\lambda)_{45}$, and $I(\lambda)_{135}$ are intensities measured at different position of the phase plates.

### 3.3. Spectropolarimetry

For the long-slit spectroscopy mode, the primary processing of observed images was performed with the specialized software packages in the IDL environment developed in the SAO RAS (Afanasiev and Amirkhanyan, 2012). The slit was oriented along the comet's velocity vector for both sets of observations of the comet in spectral mode. The data reduction included the bias subtraction, the flat field correction, geometrical correction along the slit, correction of the spectral line curvature, the sky background subtraction, spectral sensitivity of the instrument, the spectral wavelength calibration, the presentation of data with uniform scale spacing along the wavelengths, extraction of the spectra from the images, and at last, the calculation of the Stokes parameters. A large advantage of the design of the instrument SCORPIO-2 is that it does not introduce any significant instrumental polarization along the slit height.

In the case of spectropolarimetry, a dual-beam configuration of the instrument can lead to strong geometric distortions of the obtained images, which can reach up to 10–15% in the observations covering a wide spectral range. In order to obtain a high accuracy of the Stokes parameters, the scale distortions have to be carefully calibrated and exact algorithms of coordinate conversion have to be applied. Examples of images at various stages of data reduction, obtained with the single Wollaston prism, are shown in Fig. 1. We registered two distributions of the intensity of the $I_o(\lambda)$ and $I_e(\lambda)$ rays with wavelength (see Fig. 1a). All the raw images (comet, flat field, neon) were corrected for the geometric distortions using the IDL environment WARP_TRI procedure. Fig. 1b shows the spectra after the procedure of compensation of the spatial curvature of spectral lines along the slit height. The corrections for the sensitivity heterogeneity, i.e. the flat-field procedure, are also done. To estimate the level of sky background, we obtained the night sky spectrum (Fig. 1c) at similar positions of the analyzer, which was subtracted from the spectra of the comet. The quality of the night sky spectrum subtraction is especially important for the polarization observations of faint extended objects, such as cometary coma. The ultimate spectra of the comet are shown in Fig. 1d.

The wavelength scale calibration was performed by a standard way: automatic line identification, two-dimensional approximation of the dispersion curve by the third-order polynomial, quadratic smoothing of the polynomial coefficients along the slit height, and image linearization. As an example, images of intensity spectrum with numerous cometary emissions (top image) and linear polarization (bottom image) of comet Garradd are illustrated in Fig. 2.

Since the procedure of decomposition (of the observed light as a function of wavelength along the slit) of the comet increases the statistical errors, the obtained spectra were integrated along the slit. For the estimates of the Stokes parameters ($I_\lambda$, $Q_\lambda$, and $U_\lambda$) and consequently the linear polarization ($P_\lambda$) and polarization angle ($\theta_\lambda$), we calculated the flux for each spectral channel according to expression $F_\lambda=(I_o(\lambda)-I_e(\lambda))/(I_o(\lambda)+I_e(\lambda))$, where $I_o$ and $I_e$ are the intensities measured at different angles of the $\lambda/2$ phase plate (0°, 45°, 22.5°, and 67.5°). After that we found the Stokes parameters as $Q_\lambda=\frac{1}{2}(F_\lambda(0°)-F_\lambda(45°))$, $U_\lambda=\frac{1}{2}(F_\lambda(22.5°)-F_\lambda(67.5°))$, $P$ and $\theta$ as $P_\lambda = \sqrt{(Q_\lambda^2 + U_\lambda^2)}$ and $\theta_\lambda = 1/2\ arctg(U_\lambda/Q_\lambda)$. In further analysis, the polarization parameters ($I_\lambda$, $Q_\lambda$, $U_\lambda$, $P_\lambda$, $\theta_\lambda$) were robustly estimated as average for all cycles of measurements. The required quantities are controlled within three standard deviations from a mean (the 3-sigma criterion): values more than 3-sigma were ignored. This type of measurements gives a good estimate of statistical errors of the measured parameters. The degree of linear polarization of the comet was obtained as a function of wavelength in the 3 × 10 arcsec area from the center of the comet (see Section 4.3).

### 3.4. Analysis of errors

The accuracy of measurements of polarization parameters is limited by count statistics from object and sky background as well as errors of instrumental polarization. Several standard stars with zero and large polarization degree were observed during sets of our observations of comet Garradd. The linear and circular instrumental polarizations were stable, and their errors did not exceed 0.05% and 0.005%, respectively. Errors from a point-spread function are taken into account in directly from observations of polarization standards. Thus, the count statistics is the main contributor in the error of polarization degree.

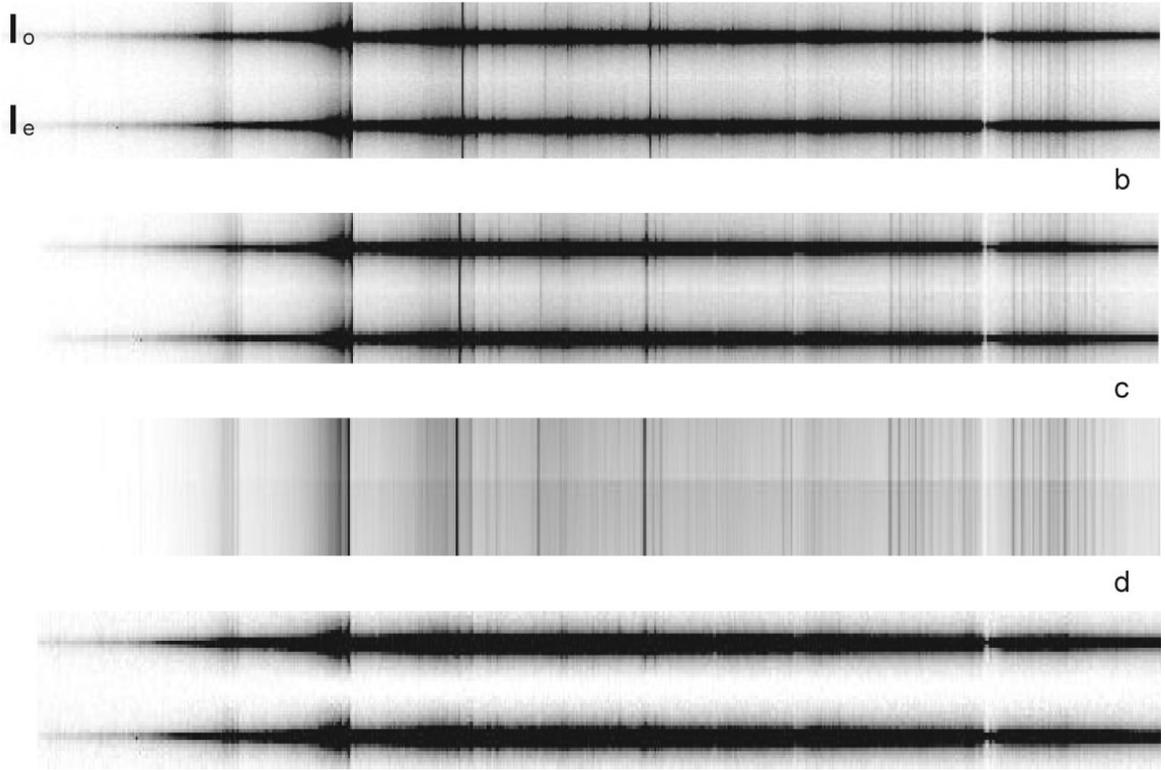

**Fig. 1.** Step by step processing of the comet Garradd spectra derived with the WOLL-1 on February 2.086, 2012: a – raw spectra $I_o(\lambda)$ and $I_e(\lambda)$; b – the spectra corrected for the geometric distortions; c – sky spectrum; d – the cometary spectra after the subtraction of the night-sky spectrum.

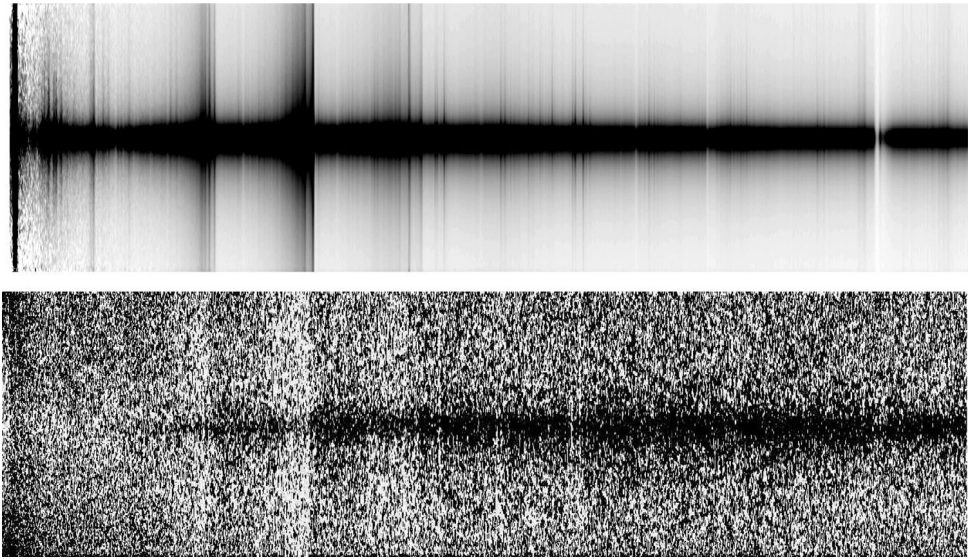

**Fig. 2.** The long-slit spectra of comet Garradd derived on February 2.099, 2012. Intensity (top panel) and linear polarization (bottom panel) are displayed.

Statistics of photons incident on the radiation detector is described by the Poisson distribution. The transition from the observations at single-channel and two-channel polarimetric devices with a fixed aperture for which it was believed that the registered pulses obey the Poisson law (see Shakhovskoj and Efimov (1972)), to the observations with a CCD camera showed that counts statistics may significantly deviate from a Poisson distribution. The distortion of statistics may be due to instrumental effects, namely, a heterogeneity sensitivity of the matrix, readout noise, non-linearity of the receiver, impulse noise (cosmic rays), etc. In order to detect deviations from a Poisson distribution, we tested our measurements of circular polarization in comet Garradd obtained on February 14, 2012. For this, we built a dependence of the index of dispersion, which was equal to the ratio of the dispersion (rms$^2$, where rms is the root mean square) to the average value (Cox and Lewis, 1966), on the Stokes parameter V. In the case of a Poisson distribution, the dispersion index must be equal to unity.

Fig. 3 shows that counts statistics for our sample data differs from the Poisson distribution. Growth of the index of dispersion at small intensities is associated with the readout noise, whereas increase of this index at large intensity is explained by non-ideality correcting of the flat field. Even in the case of the best signal/noise ratio for the observational data, the index of dispersion differs from the Poisson one tripled. Hence, when we analyze data derived on panoramic radiation detectors (CCD, etc), we have to use other methods for estimating the accuracy of the degree of polarization. For our data, we used the method of robust estimators (histogram analysis) which is not sensitive to various deviations and irregularities in the sample (Maronna et al., 2006). Such a technique allows one to carry out reduction of data correctly, without displacing the statistical evaluation. For example, the removal of cosmic rays in the images by using different methods of smoothing leads to appearance of great instrumental polarization, up to 5%, which is completely eliminated when histogram analysis is used.

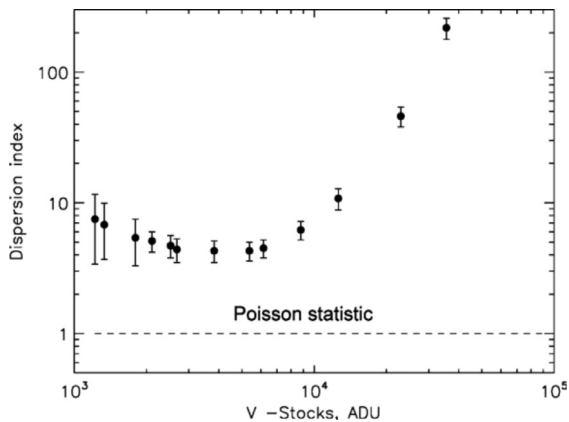

**Fig. 3.** The dispersion index for the degree of circular polarization of comet Garradd vs the signal strength (in units of ADU) obtained for the observations on February 14, 2012.

Fig. 4 (a and b) shows the histograms of the degree of circular polarization obtained by us for comet Garradd in the annular aperture of width 7″ for the two sets of observations. These figures show that in the first approximation the distribution of our estimates of circular polarization is close to the Gaussian statistics, which, as shown by Serkowski (1958) and Clark et al. (1983), is more plausible for the statistical description of normalized Stokes parameters than the Poisson statistics.

## 4. Analysis of observational data

### 4.1. Intensity images

#### 4.1.1. Coma morphology

The direct images of comet Garradd along with relative isophotes are shown for three dates of observations in Fig. 5. Extensive coma with highly concentrated material in the near-nucleus area of the comet was observed during whole observational period. More confident, we can talk about the dust only in the case of observations on February 14 (middle plot), when the comet was observed through the narrow-band filter RC, as the broadband observations of the comet on February 2 and April 14 included a gas component. The weak sunward-antisunward asymmetry of the coma is observed on February 2 and 14, whereas on April 14 the asymmetry is in a perpendicular direction to the solar direction (Fig. 6).

The treated intensity images derived with the broadband filters (Figs. 6a, c) as well as with narrowband cometary continuum filter RC (Fig. 6b) showed two bright structures in the coma which we labeled as J1 and J2. In projection on the sky, the fainter structure J1 is directed toward the Sun, whereas the bright structure J2 is oriented in the antisolar direction. The impact of these structures on the shape of isophotes is seen in Fig. 5: due to the presence of structure J1, the external isophotes are elongated in the solar direction on February 2 and 14 (upper and middle panels) and perpendicular to the solar-antisolar direction on April 14 (bottom panel). At the same time, the near-nucleus isophotes (middle and bottom panels in Fig. 5) are compressed opposite to the Sun, i.e. where the bright structure J2 is located. On February 2.099, 2012, the position angles of structures J1 and J2 were 119° and 298°, respectively; on February 14.076, 121° and 304°; on April 14.843, 278° and 73° Hadamcik et al. (2014) indicates structures (fan-shaped fine jets) at position angles from 55° to 120° which observed in the solar direction during February. Comparison of our images and the position angles of features in the coma with the available data in the literature, including amateur observations (e.g., http://www.aerith.net/comet/catalog/2009P1/pictures.html; http://kometen.fgvds.de/pix/2009P1.htm), showed that we observed two tails in comet: dust tail which complies with the jet J1 and gas tail corresponding to the jet J2. As we can see in Fig. 6b, the images obtained through the cometary continuum filter RC ($\lambda$0.684/0.090 μm) show a strong dust tail (J1) oriented to the Sun and significantly weaker gas tail (J2) than that obtained in the broadband filters (a and c images). However, the red continuum (RC) filter transmits quite a lot of emission radiation.

#### 4.1.2. Rotation period of the nucleus

To analyze the angular shifts of the structures in the cometary coma (which allows us to estimate the rotation period of the comet), the cross-correlation method was used (Ivanova et al., 2012). First, all the images were transformed from rectangular to polar coordinates with a center corresponding to the optocenter of the comet. The radial distance from the nucleus and the azimuth angle counted off counterclockwise from the north direction were taken as the polar distance and the polar angle, respectively. Also, before the coordinate system transformation, all the images were oriented in the same way, according to the north direction and the east direction. To avoid probable erroneous results, we removed the low-frequency trend in the images of the comet in polar coordinates that were not subject to digital processing. The structures selected in such a way were further used to evaluate the rotation period of the cometary nucleus.

We used the structure J2 in Fig. 6 to estimate the rotation period of the nucleus of comet Garradd. Fig. 7a shows the structures superposed with an accounting for the value of the rotation period determined with the cross-correlation method. Fig. 7b presents the general pattern of shifts for the selected structure and the approximation curve serving as a basis for determining the rotation period for two sets of observations of the comet. The difference in position angles of J2 between February 2.099 and February 14.026 was used to tentatively measure the period of rotation of the nucleus. In two weeks, the shape of structure J2 is not too differed. The rotation period estimated for comet Garradd is 11.1 ± 0.8 hours. The result obtained is in satisfactory agreement with the value of rotation period of other authors (Farnham et al., 2012), which is equal to 10.4 ± 0.05 hours.

### 4.2. Maps of circular polarization

Circular polarization maps derived on February 14.092 and April 21.795, 2012 in the continuum narrowband RC filter are presented in Fig. 8 (left panel). In plots on the right, we show correspond in one-dimensional cuts of circular polarization maps along the direction toward the Sun and perpendicular to this direction. As one can see, in both dates, circular polarization is left-handed (negative). We use a definition of left-handed circular polarization according to (Mishchenko et al., 2002): circular

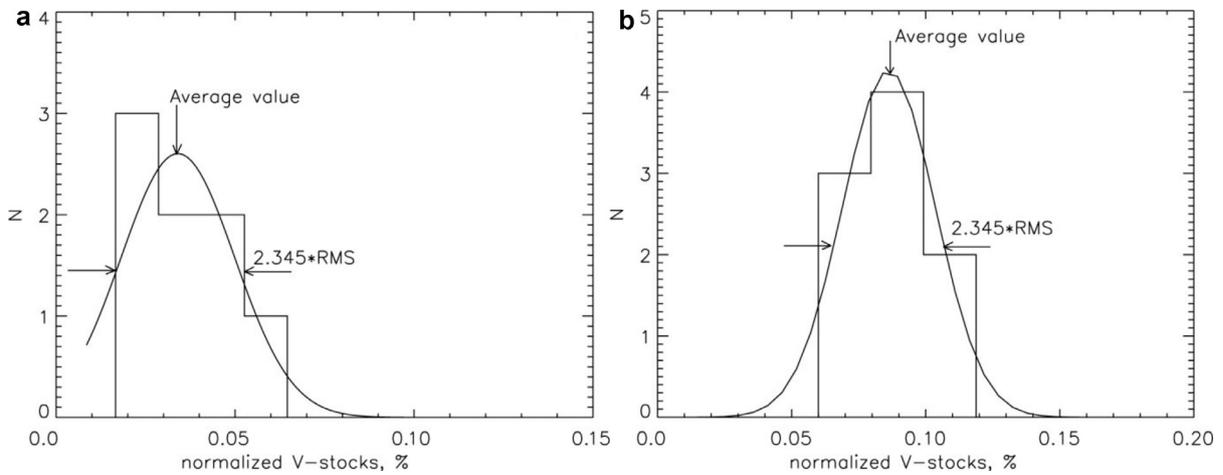

**Fig. 4.** Histogram estimates of circular polarization in the 20-px-wide (~7″ at the comet) circular apertures calculated for observations of comet Garradd on February 14, 2012 (a) and April 21, 2012 (b), respectively.

**Table 2** Linear polarization of comet Garradd from spectropolarimetric data.

| $\lambda_o$ (μm) | FWHM (μm) | Notations | $P \pm \sigma P$ (%) | |
|---|---|---|---|---|
| | | | Feb. 02.086 $\alpha=35.9°$ | Apr. 14.864 $\alpha=27.4°$ |
| 0.445[a] | 0.062 | BC | 4.7±0.2 | 3.4±0.2 |
| 0.465[b] | 0.050 | g-sdss | 4.5±0.2 | 3.0±0.2 |
| 0.514[a] | 0.086 | $C_2$ | 3.1±0.2 | 2.8±0.2 |
| 0.526[a] | 0.057 | GC | 4.3±0.2 | 3.3±0.2 |
| 0.551[b] | 0.088 | V | 3.8±0.2 | 2.8±0.2 |
| 0.580[c] | – | – | 4.0±0.2 | 3.0±0.2 |
| 0.684[b] | 0.090 | RC | 4.8±0.2 | 3.9±0.2 |
| 0.740[c] | – | – | – | 4.0±0.2 |
| 0.830[c] | 0.062 | RC | 5.0±0.2 | – |

(a) – central wavelength of HB narrowband comet filters (Farnham et al., 2000); (b) – central wavelength of filters used during our observations at the 6-m telescope of SAO RAS; (c) – central wavelength used for calculation of spectral gradient of polarization.

polarization is left-handed (negative) when the electric vector rotates counterclockwise, as viewed by an observer looking in the direction of the light propagation.

On February 14.092 (top right plot), the mean value of the degree of circular polarization averaged over the area approximately 20,000 km × 20,000 km in size was about $(-0.06 \pm 0.02)$%. Distribution of circular polarization over the coma was symmetrical and almost constant up to $\sim 3 \times 10^4$ km from the comet nucleus. On April 21.795 (bottom right plot), the distribution of circular polarization over the coma was less symmetrical. The degree of circular polarization in the near-nucleus area was about $(-0.12 \pm 0.02)$% and significantly increased (in absolute value), up to $-(0.4 \div 0.5)$% in the outer coma at the distance from the nucleus $\sim 25,000$ km. There are little differences in cuts in the direction to the Sun and perpendicularly to it, about 0.05–0.1% in magnitude.

### 4.3. Spectropolarimetry

The distribution of intensity and the degree of linear polarization of comet Garradd, derived on February 2.086 and April 14.864, 2012, as a function of wavelength are shown in Figs. 9 and 10, respectively. In Table 2, we presented the values of linear polarization and their errors, obtained from spectropolarimetric observations of comet Garradd for selected wavelengths which correspond to the central wavelength of the narrow- and broadband filters used for observations and in several selected wavelengths for determination of spectropolarimetric gradient. The maximum error in the degree of linear polarization in emissions and continuum did not exceed 0.2% for each date of observations.

The molecular emission bands in intensity spectrum (Figs. 9 and 10, top panels) correlate with the depressions in polarization spectrum (Figs. 9 and 10, bottom panels). It is seen that the polarization in continuum slightly increases with the wavelength (see Table 2). Actually, on February 2.086, the degree of linear polarization at phase angle 35.9° was about 4% at 0.58 μm increasing to 5% at 0.83 μm. On April 14.864, when the phase angle was 27.4°, observations showed a polarization of 3% at 0.58 μm increasing to 4% at 0.74 μm. These selected wavelengths for continuum are free of gas contamination. The spectral gradient of polarization (slope), $\Delta P/\Delta \lambda = [P(\lambda_2) - P(\lambda_1)]/(\lambda_2 - \lambda_1)$, is $+4 \pm 0.8$%/μm and $+6.2 \pm 1.3$%/μm in both dates, respectively.

The observed degree of polarization at wavelengths corresponding to the emission bands is smaller than that in the nearby continuous spectrum. As can be seen in Figs. 9 and 10 and Table 2, the degree of polarization of the emission $C_2$ ($\lambda 0.514$ μm) on February 2 and April 14 was about 2.2% and 1.5%, respectively, whereas polarization degree in the nearby continuum ($\lambda 0.4845$ μm) was about 4% and 3% in both dates. This effect can be also seen in Fig. 11, which shows the spatial profiles of brightness (a, b) and linear polarization (c, d) on February 2.061, 2012 in the continuum, within the wavelength domain (there after $\Delta\lambda$) ($\Delta\lambda = 0.67 - 0.68$ μm) and emission band $C_2$ ($\Delta\lambda = 0.49 - 0.52$ μm) along the slit. According to the polarization profiles in Fig. 11, the degree of polarization in the emission band $C_2$ ($\Delta v = 0$, $\Delta\lambda = 0.4 - 0.52$ μm) was approximately 4%, 3%, and 2.5% at about 6000, 20,000, and 40,000 km from the optocenter. The polarization observed for this sequence of the $C_2$ molecule is the total polarization from emission and continuum in this sequence. At about the same distances, the degree of linear polarization in the continuum was about 5%, 4%, and 3%, respectively.

As one can see in Table 2, the values of polarization obtained for selected wavelength, which correspond the central wavelength of the broadband g-sdss and V filters, are slightly (at the level of 1–2$\sigma$) but systematically smaller as compared to those for the nearby continuum filters BC and GC. Due to the small difference in the values of polarization and relatively large errors in the polarization, the eventual gas contamination in the broadband g-sdss and V filters was not taken into account.

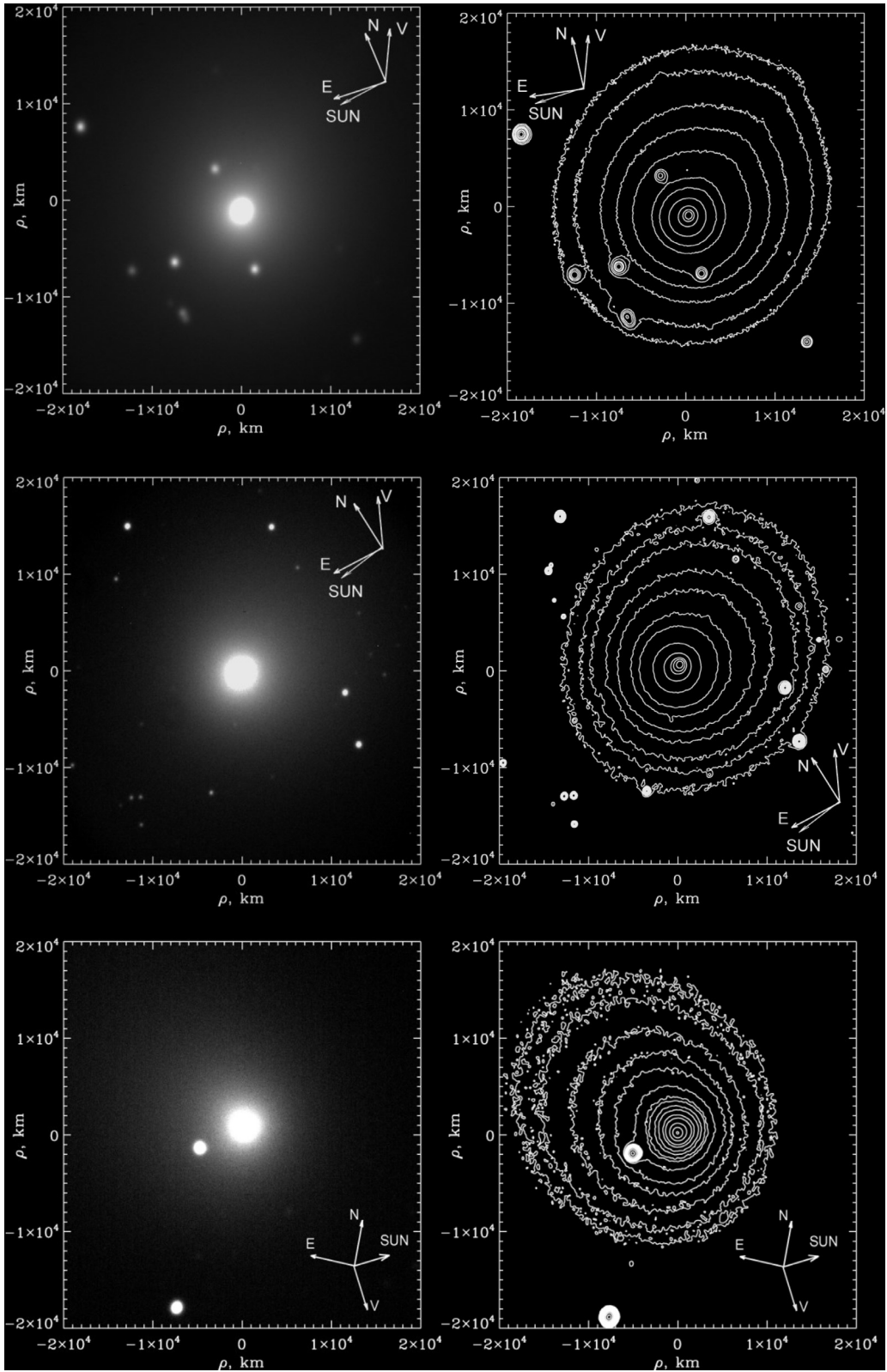

**Fig. 5.** Images of comet Garradd (left panel) and intensity map with relative isophotes differing by a factor √2 (right panel): the top plot represents the image obtained through the broad-band g-sdss filter on February 2.099, 2012; the middle plot – through the narrow-band cometary continuum filter RC λ6840 Å, February 14.076, 2012; and the bottom plot – through the broad-band V filter, April 14.843, 2012. The arrows show the directions to the Sun, North (N), East (E), and velocity vector of the comet (V).

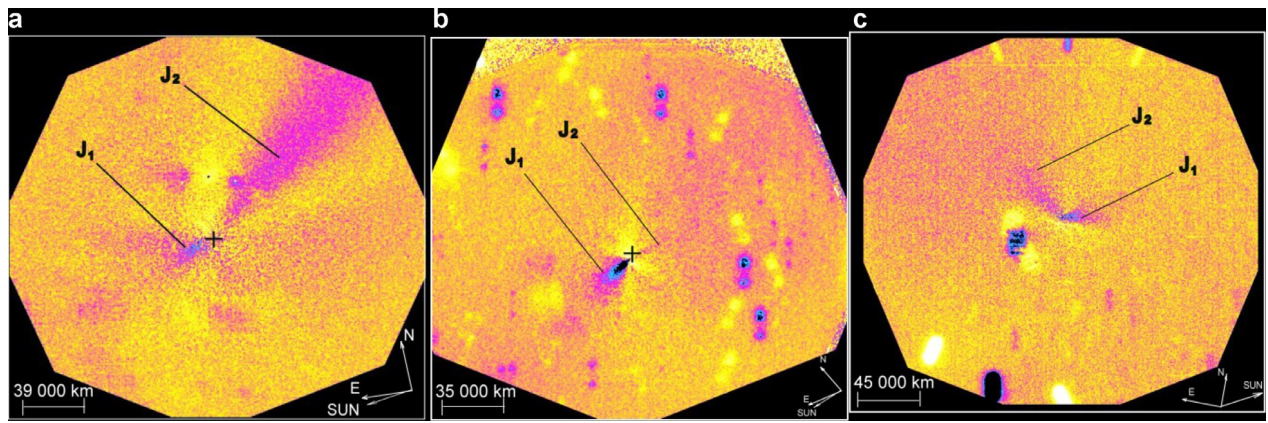

**Fig. 6.** Intensity images of comet Garradd treated by digital filters: (a) – February 2.099, 2012, g-sdss filter; (b) – February 14.076, 2012, the narrowband cometary filter RC (λ0.684/0.090 μm); (c) – April 14.843, 2012, V filter. Blurred detail on the left is a star (for comparison, see Fig. 5, bottom panel).

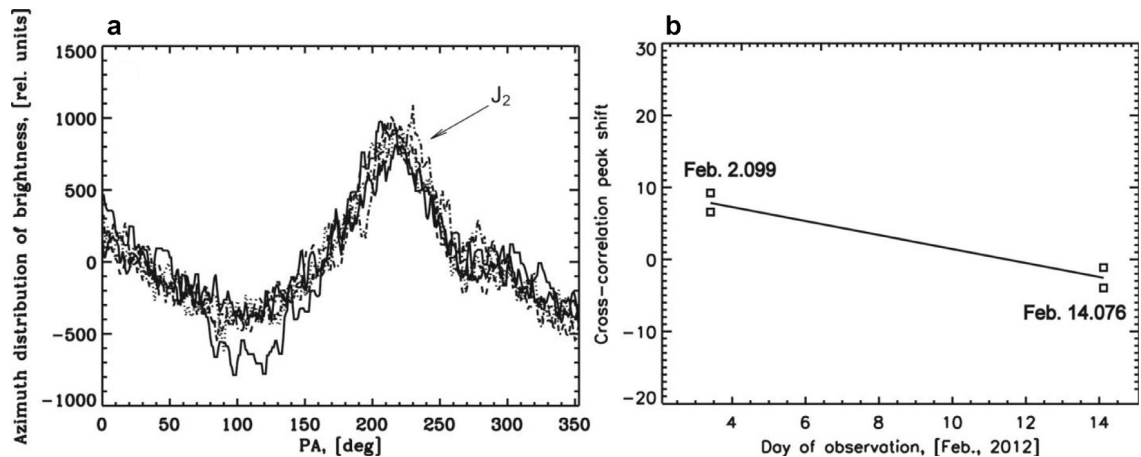

**Fig. 7.** The structure J2 in the coma of comet Garradd after subtracting from the original images (presented in a polar coordinate system) the smoothed radial profile of these images (a). Displacement of the structure J2 depending on the date of observation and approximation curve to calculate the rotation period of comet nucleus (b). Square symbols correspond to the identified structures for the two periods of observations of the comet.

### 4.4. Spectroscopy: molecular emissions in cometary coma

Obtaining cometary spectra was not the main goal of our work. Nevertheless, we used the results of spectropolarimetry observations (see, e.g., Fig. 2) to derive the distribution of energy in optical wavelengths and identify emission features in the spectra of comet Garradd (Fig. 12). To isolate the emission spectrum, we compensated the continuum of the comet by subtracting the high-resolution solar spectrum (Kurucz et al., 1984) from the cometary one. The solar spectrum was reduced to the spectral resolution of the cometary spectrum by the Gaussian convolution having the corresponding half-width of the profile (FWHM). The solar spectrum was shifted in such a way that its level was as close as possible to the lower boundary of the cometary spectrum in the spectral windows, where the continuum dominated. For the first period (February 2.086, 2012), we separated atmospheric emissions of OH and the night sky. For the identification of atmospheric emissions of OH, the observational data of Osterbrock et al. (1996) were used.

To identify the observed emission features, we calculated theoretical spectra of the molecules that had been already recorded in cometary spectra. The intensities of individual rotation lines were calculated under the thermodynamic equilibrium assumption. A vibration and rotation energy levels were populated according to the Boltzmann distribution and defined by rotational and vibrational temperatures. Although the thermodynamic equilibrium is not realistic for the case of cometary coma, the thermodynamic equilibrium approach has been successfully applied so far for the identification of molecular emissions in cometary spectra. To compare the theoretical and observed spectrum, the latter was corrected for the Doppler effect resulted from the motion of the comet with respect to the Earth.

CN: The ($\Delta v=+1$, $\Delta v=0$) vibration bands of the violet system of CN attributed to the electronic transition $B^2\Sigma^+–X^2\Sigma^+$ were detected in the blue region of the spectra. Its identification was fulfilled with the program LIFBASE (Luque and Crosley, 1999) developed for the purpose of calculation of the electronic spectra of some diatomic molecules. Since the spectral lines attributed to the rotational transitions were located close to each other on the wavelength scale of our spectrum, the moderate spectral resolution did not allow us to resolve them.

$C_3$: We detected some emission features of the $C_3$. The structure of the band was difficult for modeling because it strongly depended on the initial conditions. For identification, we used data from literature (Dobrovolsky, 1966; Gausset et al., 1965).

$CO^+$: In our spectra, we detected some weak features due to the emission. These features belonged to the head of the 2-0, 1-0, 1-1, and 0-0 bands of the $A^2\Sigma–X^2\Pi$ electronic transition. The theoretical spectrum of this molecule was calculated based on the database containing molecules that had been observed in the astronomical spectra (Kim, 1994).

CH: The emission features of the $A^2\Pi–X^2\Delta$ system of CH molecule were confidently identified in the blue region. We sep-arated emission lines of the (0–0) vibrational transition. For the identification we used LIFBASE package (Luque and Crosley, 1999).

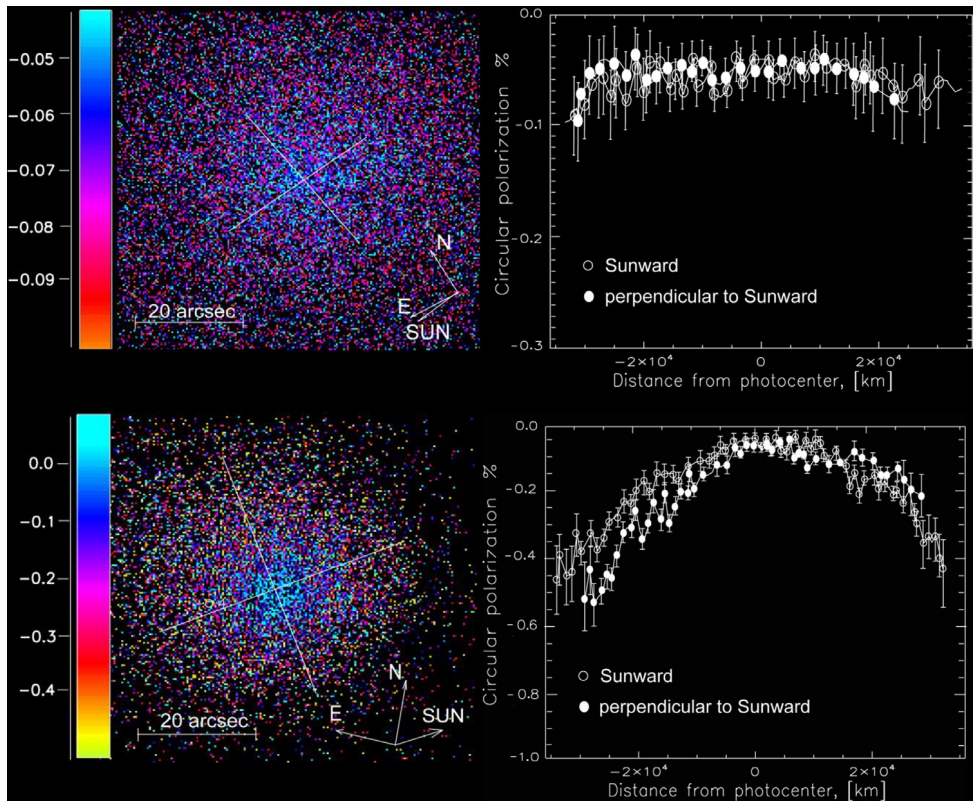

**Fig. 8.** Distribution of circular polarization in the coma of comet Garradd (left panel) and its variations with the distance from the nucleus along cuts to the sunward direction and perpendicularly to it (right panel). The top row: February 14.092, 1 arcsec= 1008 km at the comet; the bottom row: April 21.795, 1 arcsec= 1422 km. The arrows show the direction to the Sun, North, and East.

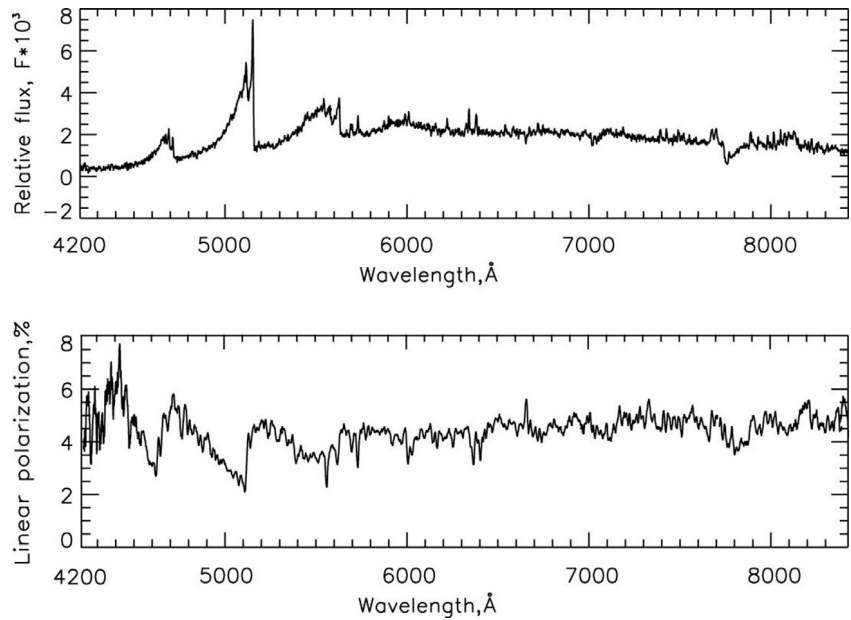

**Fig. 9.** The long-slit spectra of comet Garradd derived at phase angle 35.9° on February 2.086, 2012. The top and bottom panels display the integral intensity and the degree of linear polarization as a function of wavelength in the 3 × 10 arcsec (3329 × 11,097 km) area around the center of the comet. The level of flux for wavelengths less than 0.46 μm is very low therefore polarization at these wavelengths is an artifact.

$C_2$: We detected three vibrational band systems of $C_2$ ($\Delta v=+1$, $\Delta v=0$ and $\Delta v=-1$) to the electronic transition $A^3\Pi_g$–$X^3\Pi_u$, so called the Swan system. The theoretical spectrum was calculated using the line list derived from the laboratory measurements made by Phillips and Davis (1968). To compare the observed and calculated spectra, the latter was degraded to the spectral resolution of the former by the convolution with a Gaussian profile with a FWHM equal to the spectral resolution of the observed spectrum.

$NH_2$: The numerous weak emissions of $NH_2$ were dispersed within the same spectral window where the emission bands of $C_2$

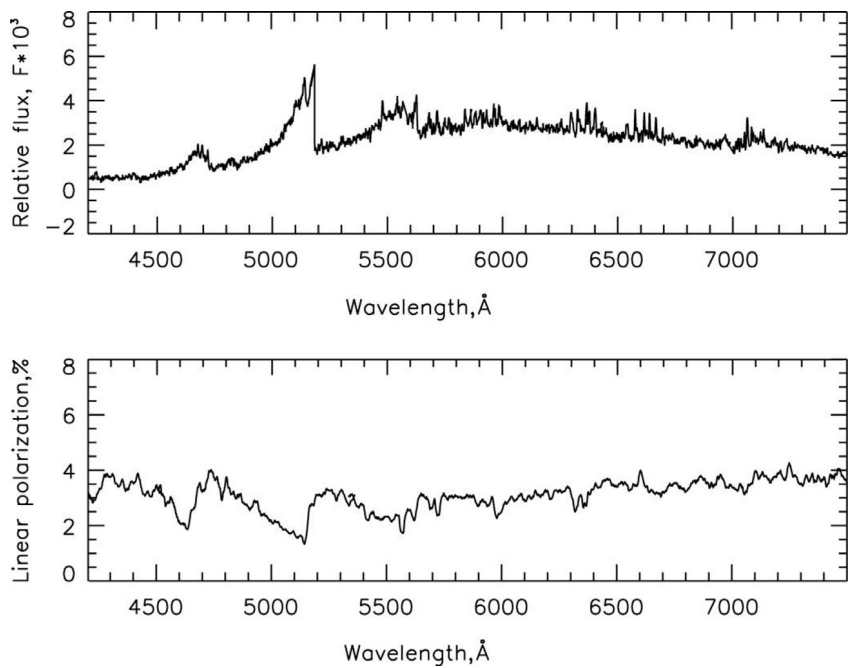

**Fig. 10.** The long-slit spectra of comet Garradd derived at phase angle 27.4° on April 14.864, 2012. The area measured around the optocenter of the comet is 3 × 10 arcsec (3916 × 13055 km). The notations are the same as in Fig. 9.

predominate. The spectrum of $NH_2$ was irregular and belonged to electronic transition $A^2A^1$–$X^2B^1$. We used the results of the laboratory measurements of Dressler and Ramsay (1959) for identify these features. The comparison between the calculated spectrum of the $NH_2$ molecule and the observed spectrum resulted in the identification of the emissions belonging to the (0,15,0), (0,14,0), (0,13,0), (0,12,0), (0,11,0), (0,10,0), (0,9,0), (0,8,0), (0,7,0), (0,6,0), (0,5,0), and (0,4,0) vibrational bands of this molecule.

$H_2O^+$: In the spectral range, where the bands of the $NH_2$ molecule were identified, we also detected emissions of $H_2O^+$. We identified emission lines of the (0,6,0), (0,8,0), (0,7,0), (0,9,0), and (0, 10.0) bands. The theoretical spectrum was calculated using data of Lew (1976).

## 5. Discussion

### 5.1. Activity and coma morphology

We observed comet Garradd at relatively small heliocentric distances, from 1.65 to 2.23 au This allowed us to obtain the images of the comet with a sufficiently high spatial resolution and analyze different features of the coma. The photometric images derived with the broadband filters (Fig. 6a, c) as well as with narrowband cometary continuum filter RC (Fig. 6b) showed two bright structures in the coma (J1 and J2). The influence of these structures on the shape of isophotes is clearly seen in Fig. 5: due to the presence of feature J1, the external isophotes are elongated, whereas the near-nucleus isophotes are compressed due to the bright structure J2. Comparison of the structures visible in all dates of our observations with available images of comet Garradd obtained by other observers shows that we revealed two tails in the comet: dust tail which coincides with the feature J1 and gas tail corresponding to the direction of structure J2. The flows of dust and gas emanating from the nucleus each form their own distinct tail, pointing different directions, when viewed from our prospective. On February 2 and 14, they were located in nearly opposite directions (dust tail in the sunward direction and gas tail in an anti-sunward direction), but on April 14 they were in nearly perpendicular directions (Fig. 6c).

Outflow of dust and gas from the comet's nucleus and the existence of the dust and gas tails for a long time are caused by activity of comet Garradd which was observed both before and after perihelion. Approximately at the same distances at which our observations were carried out, but in the pre-perihelion passage, similar activity with bright structures in the coma and gas tail was also observed in the comet (Ivanova et al., 2014; Bodewits et al., 2014). Some asymmetric increasing activity of comet Garradd before and after perihelion is seen, for example, on light curve and $Af\rho$ data (which correspond to the dust productivity in the comet) presented on M. Kidger's website.[1] The $H_2O$ and CO production rates also increased around the perihelion (Bodewits et al., 2014; Feaga et al., 2014 and references therein). Such asymmetric increase in activity of the comet around perihelion passage may be caused by seasonal effects and/or properties inherent for dynamically new comet Garradd. Feaga et al. (2014) suggested the existence of two separate active regions on different hemispheres which are activated at different times depending on the orientation of the rotational pole of the nucleus to the Sun.

The revealed features in the cometary coma were being observed for over three months that allowed us to find the period of rotation of the nucleus using the cross-correlation method. The value of rotation period is 11.1 ± 0.8 h. The result obtained is in satisfactory agreement with the value of rotation period found by other authors (Farnham et al., 2012), which is equal to 10.4 ± 0.05 h. The small difference in the obtained values of the rotation period can be explained by different methods of its calculation and different periods of observations of the comet selected for rotation period estimation. Rotation of comet nucleus can be complex (see, e.g., Festou et al. (1987)) and therefore any additional data are always useful. On the other hand, our result confirmed the location of observed features in the intensity images.

### 5.2. Spectroscopy: emissions and continuum

From spectroscopic observations of comet Garradd in optical wavelengths range 0.36–0.80 μm (February 2 and April 14,

---

[1] http://www.observadores-cometas.com/cometas/2009p1/afrho.html.

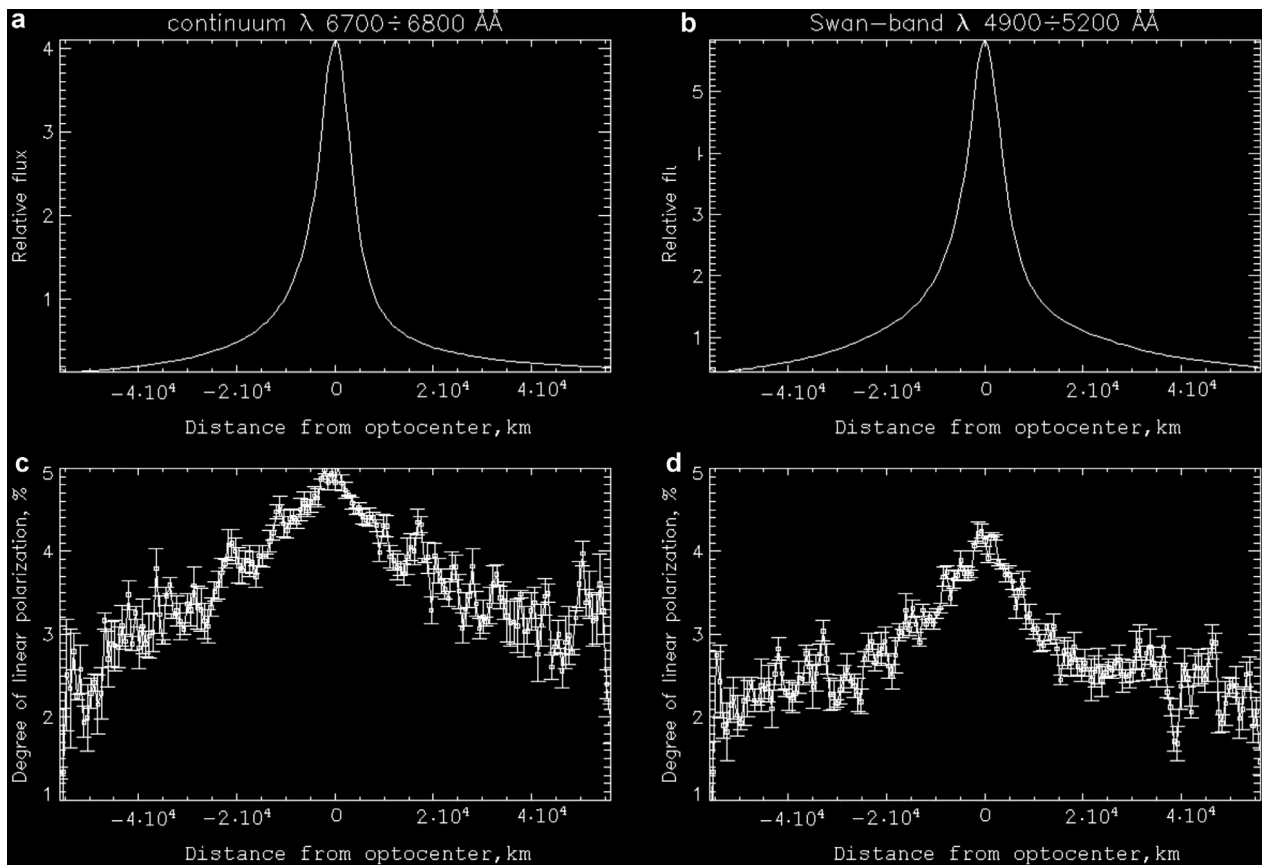

**Fig. 11.** A comparison of the spatial profiles of brightness (a, b) and linear polarization (c, d) in the continuum ($\Delta\lambda = 0.67 - 0.68\,\mu m$) and emission band $C_2$ ($\Delta\lambda = 0.49 - 0.52\,\mu m$), respectively. These profiles are obtained for observations on February 2.086, 2012 at phase angle of 35.9°.

2012), we have revealed features belonging to the $C_2$, $C_3$, CN, CH, $NH_2$ molecules as well as $CO^+$ and $H_2O^+$ ions (Fig. 12). The $C_2$ Swan bands ($\Delta \nu=-1$, $\Delta \nu=0$, and $\Delta \nu=+1$ sequences) were the strongest in the spectrum for both dates. Also, the $NH_2$ emissions were quite strong in the coma. Except for $CH^+$ band, all emissions identified in our spectra were also detected in the spectra obtained on March 20–21, 2012 by Shubina et al. (2014). Lack of the $CH^+$ band may be explained by a higher resolution of the latter. According to Paganini et al. (2012), Bodewits et al. (2014), HYPERLINK \l "bib6" Bodewits et al. (2014) , McKay et al. (2015), and Villanueva et al. (2012), the high CO abundance was detected in comet Garradd. The presence of $CO^+$ emissions detected in our spectra of the cometary coma also indicates that comet Garradd was CO rich.

The spectra (see Figs. 2, 9, 10, top panels) also indicate the high level of continuum formed by scattered sunlight from cometary dust grains in the coma. Our data confirm conclusion by Bodewits et al. (2014) and Feaga et al. (2014) that comet Garradd was dust-rich one, and its activity was complex and changed significantly over time.

### 5.3. Linear polarization in continuum

The significant dust production in the coma is also confirmed by the measurements of linear polarization of comet Garradd. Fig. 13 shows the measured linear polarization of comet Garradd according to data from Das et al. (2013), Kiselev et al. (2013), Hadamcik et al. (2014), and this work. The curves 1 and 2 in the figure are the fits to all polarization data for classes of the high-$P_{max}$ and low-$P_{max}$ comets in the red, respectively, according to Kiselev et al. (2015). One can see that difference between the polarization for high-$P_{max}$ comets (curve 1) and low-$P_{max}$ comets (curve 2) within the range of phase angles from 27° to 35°, in which we observed, are small and, hence, do not provide sufficient information to classify comet Garradd according to polarization degree. The deviation of data from the fit to low-$P_{max}$ comets becomes noticeable at phase angles larger than 35°. Therefore we can assume that comet Garradd most likely belongs to the class of high-$P_{max}$ comets.

In addition, the degree of linear polarization found from spectropolarimetry at phase angle 35.9° (Feb. 2.086) is in a good agreement with other available data (Hadamcik et al., 2014), whereas it is noticeably higher at phase angle 27.4° (Apr. 14.864) (see Fig. 13). The reason for this may be a coincidence of projection of the spectrograph slit with an area of the coma with a high degree of polarization (e.g., the jet-like features with freshly ejected dust particles). Different distribution of polarization over the coma was observed in a number of comets (see references in (Kiselev et al., 2015).

Our measurements show (Fig. 11) that on February 2 the degree of linear polarization in the continuum between 0.67 and 0.68 $\mu m$ was ~5% in the central region, up to ~6000 km distance; at about 20,000 km distance, it was approximately 4% and decreased to ~3% at ~40,000 km. The errors in the degree of polarization are 0.2%. We have compared our results with those obtained by other authors at close phase angles. According to imaging polarimetry of Garradd by Hadamcik et al. on February 18 – 20, 2012 at $\alpha \approx 34.5°$, the central region, up to 7000 km distance, showed a polarization value P between 3% and 5%, whereas in the surrounding region, at about 12,000 km distance, P was approximately 3% and decreased farther away. The values of polarization obtained by both Hadamcik et al. and Das et al. have the same tendency to decrease with increasing distance from the nucleus. Our results confirm this spatial trend. Such variations of polarization in the inner and outer coma are usually observed in comets (Kiselev and Rosenbush, 2004).This may indicate the evolution of the physical properties (size distribution, composition, and structure) of dust particles ejected from the nucleus due to solar insolation and radiation pressure that was confirmed by the spatial maps of cometary albedo increasing with distance from the nucleus (Hammel et al., 1987).

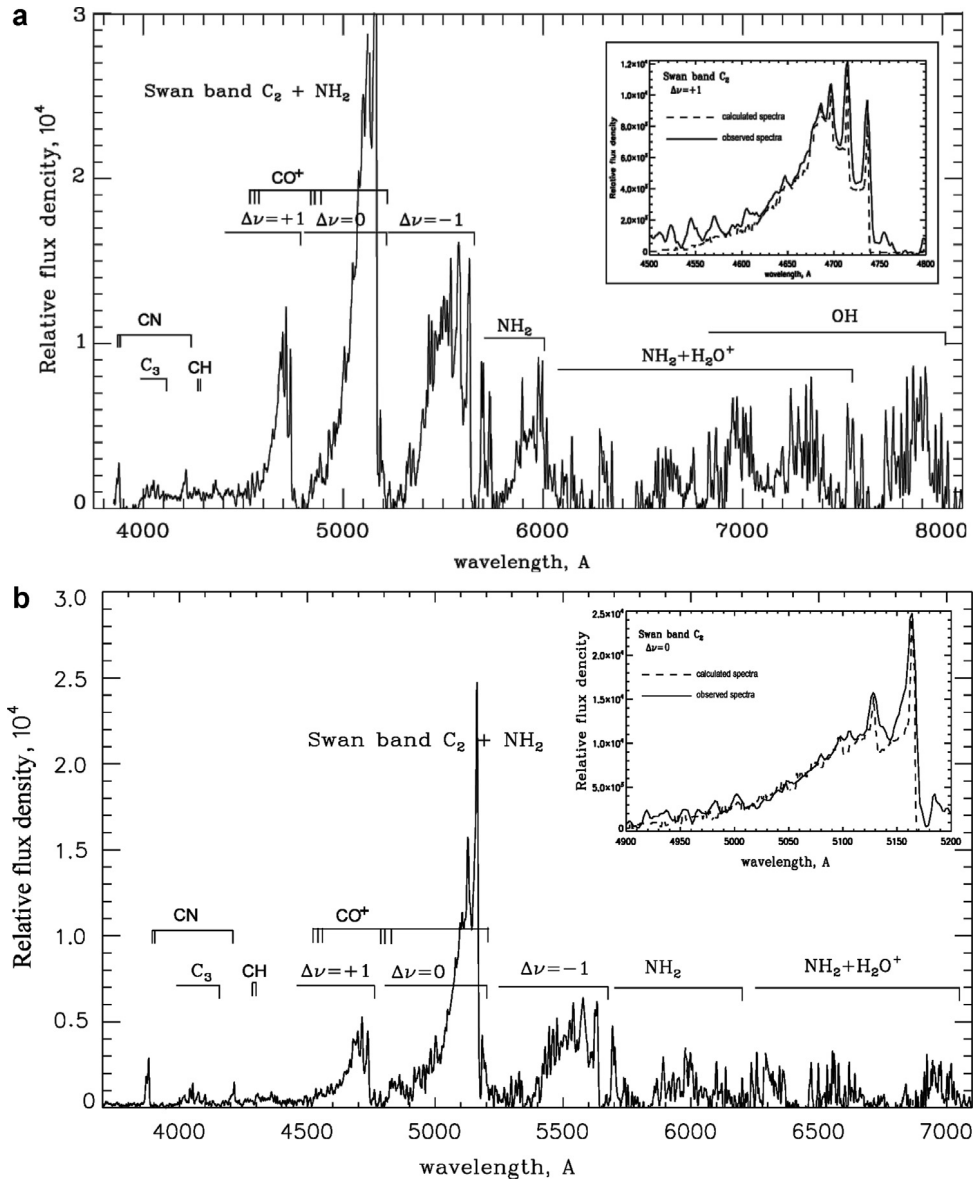

**Fig. 12.** The spectra of comet Garradd obtained on February 2.086 and April 14.864, 2012 (top and bottom panels, respectively). The sky background and reflected sunlight by cometary dust particles in the coma have been subtracted. The inset shows the observed spectrum (solid line) and the modeled spectrum (dash line) of the Swan bands $C_2$ ($\Delta \nu=+1$) (top panel) and $C_2$ ($\Delta \nu=0$) (bottom panel) as comparison.

It is known that most observations of cometary dust (Kiselev et al., 2015) shows an increase in the linear polarization with the wavelength in the visible domain for phase angles greater than about 30°. We also noted a small increase of polarization with the wavelength seen from spectropolarimetry of comet Garradd (Figs. 9 and 10).[2] On February 2.086, and April 14.864, 2012, the spectral gradient of polarization $\Delta P/\Delta \lambda$, or slope, was $+4\pm0.8\%/\mu m$ and $+6.2\pm1.3\%/\mu m$ in both dates, respectively. It is close to the mean value for dusty comets (8% per μm) at these phase angles (Kiselev et al., 2015). Moreover, whereas the degree of linear polarization slightly increases in the red domain of spectrum, the intensity slightly decreases with the wavelength. It looks like the Umov law, the effect of anticorrelation between albedo and polarization known for the regolith surfaces.

### 5.4. Contamination effect

As can be seen in Fig. 2, it is difficult to allocate the pure continuum that is free from contamination by the emission bands. Spectropolarimetric and spectrophotometric data allowed us to study the impact of emissions on polarization in the continuum and the continuum on the polarization in emissions. Furthermore, the long-slit spectra of comet Garradd allowed us also to measure the polarization in the continuum and emission bands as a function of cometocentric distance (see Fig. 11). The contamination effect by low polarization of the $\Delta \nu=-1$, $\Delta \nu=0$, and $\Delta \nu=+1$

---

[2] Note that Hadamcik et al. (2014) have not found spectral dependence of the linear polarization for comet Garradd exceeding the observational errors which varied from 0.2% to 0.8%. The cause of this discrepancy may be a higher precision of spectropolarimetric observations in comparison with imaging polarimetry.

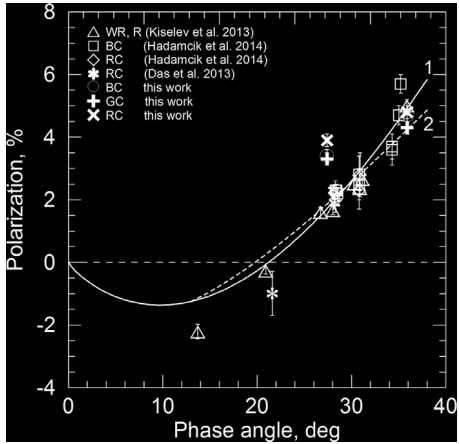

**Fig. 13.** The phase-angle dependence of linear polarization for comet Garradd according to data from Kiselev et al. (2013) (triangles), Hadamcik et al. (2014) (squares and diamonds for the BC and RC filters, respectively), Das et al. (2013) (stars), and this work (open circles, plus, and oblique cross for the BC, GC and RC filters, respectively). Curves 1 and 2 are the fits to all polarization data for high-$P_{max}$ and low-$P_{max}$ comets in the red continuum, respectively, according to Kiselev et al. (2015).

Swan-band sequences of the $C_2$ molecule in comet Garradd is very small. Actually, Figs. 9 and 10 show that the short-wavelength wing of the $\Delta v = 0$ sequence of the $C_2$ molecule extends to the domain of λ0.4845/0.0065 μm which was adopted as the blue continuum in comets. The accurate accounting of the depolarizing effect of the emission of $C_2$ on polarization in the continuum on the basis of available data cannot be made because of the low flows in the wings of the emission band, the high level of continuum, and little difference in the polarization of resonance fluorescence and continuum at these phase angles, taking into consideration that this difference is comparable with errors of observations. Other emissions in the red domain had no significant effect on the continuum polarization due to small gas-to-dust ratio (Figs. 2, 9, 10). This is confirmed by data of Das et al., (2013) who found that polarization of Garradd in the broad R filter and in the red continuum comet filter was the same. However, the depolarizing effect of continuum should be stronger for comets with a low level of continuum and strong molecular emissions (like comet Encke) and especially at large phase angles (Jockers et al., 2005). It is clear that the wide-band polarimetric data for comets in the blue and visible domains will be strongly contaminated by low polarization of the gas emissions, even for comets with the medium gas-to-dust ration.

Contrary, since the molecular emissions are observed at the continuum background, therefore there is an effect of continuum contamination on the polarization in molecular emissions. As a result, the observed polarization of the light emitted from the $C_2$ gas molecules is higher than theoretical values for resonance fluorescence mechanism. Actually, according to polarization profiles of the $C_2$ molecule, the degree of polarization in the emission band $C_2$ ($\Delta v = 0$) was approximately 4%, 3%, and 2.5% at the nucleocentric distances about 6000 km, 20,000 km and 40,000 km, respectively. Since the degree of polarization of the $C_2$ emission should not vary with distance from the nucleus, it is obvious that there is an influence of polarization in continuum on the polarization in the emission. We have estimated from Fig. 11 (a, b) that the ratio $F_{cont}/F_{C_2}$ is of the order of 4/6. On the other hand, the observed polarization in Fig. 11 (c, d) is of the order of 5% and 4% for continuum $P^{obs}_{cont}$ and emission $P^{obs}_{C_2}$, respectively. Then using relations $P^{obs}_{C_2} = P_{C_2} \times F_{C_2} + P^{obs}_{cont} \times F_{cont}$ and $F_{C_2} + F_{cont} = 1$, one can roughly estimate non-contaminated value of the polarization, $P_{C_2} = 3.3\%$ that, as expected, is below the observed polarization degree in the $C_2$ band.

The theoretical Öhman's relation $P(\alpha) = P_{90}\sin^2\alpha/(1+P_{90}\cos^2\alpha)$ for $C_2$ band polarization at phase angle $\alpha = 35.9°$ gives polarization degree of 2.5%. The value $P_{90} = 0.077$ is taken from Öhman (1939). The difference between $P_{C_2} = 3.3\%$ and expected theoretical value of 2.5% is likely due to errors in observed fluxes and polarization in continuum and $C_2$ emission band. The polarization in the $C_2$ band at the cometocentric distances about 40,000 km is the same as the theoretical value of 2.5%. At this distance the level of continuum is very small (Fig. 11). It should be noted that, unfortunately, data on the polarization in molecular emissions observed in comets are still rare (Le Borgne et al., 1988; Kiselev et al., 2005). The most of these data were obtained for comets with strong continuum and therefore are not suitable for analysis due to the effect of continuum contamination on the polarization in the molecular emissions.

### 5.5. Distribution of circular polarization over the coma

Knowledge of the characteristics of the scattered light, including linear and circular polarizations, allows to get more information on the scattering medium and, hence, to define characteristics of the scatterers more accurately. In the case of comets, linear polarization is a well-studied phenomenon, whereas the origin of circular polarization in the cometary environment is a mystery up to now. The reason for this is a very small number of comets with reliably measured non-zero circular polarization of the scattered light. On the other hand, regular mechanisms (such as multiple scattering in optically thick medium, domination of particles or materials of a specific mirror asymmetry, including homochirality, and alignment of particles) cannot explain the available observations (see, e.g., Kiselev et al. (2015) and references therein). Therefore, the measurement of circular polarization in each accessible comet is of particular interest.

Low, at level approximately from −0.06% to −0.42%, but significant left-handed circular polarization was measured in comet Garradd at distances up to $3 \times 10^4$ km from the nucleus on February 14 and April 21, respectively. On April 21, distribution of circular polarization in the coma was not symmetrical. There is also some systematic increase of the degree of circular polarization to the outer edge of the coma in the sunward direction as well as in perpendicularly to it: up to −(0.4÷0.5)% at the distance from the nucleus ~25,000 km. Increasing the value of circular polarization with the distance from the nucleus observed on April 14 is probably due to an increase of fine dust particles in the coma. According to Kidger's website (see note 1), in April the value $Af\rho$ was increased in the comet. It is interesting to note that the circular polarization in the detected features seems to be the same as in the surrounding coma (Fig. 8). In addition to comet Garradd, we have detected rather low but significant left-handed circular polarization in several other comets, such as C/2001 Q4 (NEAT), C/2011 R1(McNaught), C/2012 K1 (PanSTARRS), 2014 R1 (Borisov), C/2011L4 (PanSTARRS), 29P/Schwassmann−Wachmann 1, 18P/Tuttle. Contrary, circular polarization was absent in comets 290P/Jager, 108P/Ciffreo, and 2P/Encke (are being processed). Mechanisms of formation of cometary circular polarization are still not identified and are a subject of numerous discussions (see Kiselev et al. (2015), Kolokolova et al. (2016)).

### 6. Summary

Our observations of comet Garradd were performed in period of February 2 − April 21, 2012, when it already passed through perihelion on December 23, 2011 at 1.55 au from the Sun and was at the heliocentric distances from 1.65 to 2.23 au In that time, the phase angle was between 35.9° and 26.8°. From our observations we have derived the photometric images of the comet, spec-tra of intensity and polarization, and circular polarization images. The main results provided by our observations can be summarized as follows:

1) Two features (dust and gas tails) oriented in the solar and antisolar directions on February 2 and 14, 2012 were revealed in treated images of comet Garradd that allowed us to determine the period of rotation of the nucleus as $11.1 \pm 0.8$ h.
2) Emission bands of neutral molecules such as $C_2$, $C_3$, CN, CH, and $NH_2$ as well as $CO^+$ and $H_2O^+$ ions together with very strong continuum were identified in the spectra of the comet Garradd.
3) On February 2, the degree of linear polarization at phase angle 35.9° was about 4% at 0.58 µm increasing to 5% at 0.83 µm, while on April 14, when the phase angle was 27.4°, observations showed the polarization ~3% at 0.58 µm increasing to 4% at 0.74 µm. The errors in the degree of polarization are 0.2%. This corresponds to the spectral gradient of polarization $\Delta P/\Delta \lambda + 4 \pm 0.8\%/\mu m$ and $+6.2 \pm 1.3\%/\mu m$ in both dates, respectively.
4) Linear polarization of Garradd is consistent with that for so-called class of dust-rich or high-$P_{max}$ comets. On February 2, the degree of linear polarization in the continuum (within the range $0.67 - 0.68\,\mu m$) was ~5% in the central region up to ~6000 km from the nucleus, at about 20,000 km it was ~4% and decreased to ~3% at ~40,000 km.
5) Due to continuum domination in the spectrum of comet Garradd, the gas flux contribution in the dust continuum wavelengths range is small as seen in linear polarization. In the emission bands, the dust contribution is important and needs to be corrected to obtain the correct polarization in these bands (obtained by theory). We found that the non-contaminated (due to the continuum) the degree of polarization in the $C_2$ ($\Delta \nu = 0$) emission band is about 3.3% that is slightly higher than the theoretical value of 2.5% at phase angle 35.9°.
6) The significant left-handed (negative) circular polarization was detected at distances up to $3 \times 10^4$ km from the cometary nucleus with values from about –0.06% to –0.5% (with errors 0.02%) on February 14 and April 21, respectively. There is some systematic increase in the degree of circular polarization to the outer edge of the coma on April 21.

## Acknowledgments

The observations at the 6-m BTA telescope were carried out with the financial support of the Ministry of Education and Science of the Russian Federation (agreement No. 14.619.21.0004, project ID RFMEFI61914×0004). The authors also express appreciation to the Large Telescope Program Committee of the RAS for the possibility of implementing the program of spectropolarimetric observations at the BTA. O. Ivanova thanks the SASPRO Programme, the People Programme (Marie Curie Actions) European Union's Seventh Framework Programme under REA grant agreement No. 609427, and the Slovak Academy of Sciences for financial support. We are grateful to L. Kolokolova for helpful discussions. Also, we truly appreciate the Reviewers for very careful review of our paper, whose comments and suggestions, we believe, greatly improved the paper.